\documentclass[a4paper,preprintnumbers,floatfix,superscriptaddress,pra,twocolumn,showpacs,notitlepage]{revtex4-1}

\usepackage[utf8]{inputenc}
\usepackage[T1]{fontenc}
\usepackage[sc,osf]{mathpazo}
\usepackage{amsmath, amsthm, amssymb,amsfonts}
\usepackage{graphicx}
\usepackage{dcolumn}
\usepackage{bm}
\usepackage{bbm}
\usepackage{hyperref}
\usepackage{mathtools}
\usepackage{comment}
\usepackage{color}

\def\1{\mathbf{1}}
\def\0{\mathbf{0}}

\def\p{\mathbf{p}}
\def\E{\mathbf{E}}

\newcommand{\mean}[1]{\left\langle #1 \right\rangle}

\newcommand{\beq}{\begin{equation}}
\newcommand{\eeq}{\end{equation}}

\renewcommand{\rho}{\varrho}

\newcommand{\processnext}[1]{%
  \ifx\listfinish#1\empty\else\listact{#1}\expandafter\processnext\fi}

\newcommand{\figref}[1]{Fig.~\ref{#1}}

\newcommand{\ea}{\end{eqnarray}}
\newcommand{\ban}{\begin{eqnarray*}}
\newcommand{\ean}{\end{eqnarray*}}

{\begin{framed}\begin{small}}
{\end{small}\end{framed}}

\DeclareGraphicsExtensions{.pdf,.png,.jpg}

\makeindex

\begin{document}
\title{Polynomial Bell inequalities}

\author{Rafael Chaves}
\affiliation{Institute for Physics \& FDM, University of Freiburg, 79104 Freiburg, Germany}
\affiliation{Institute for Theoretical Physics, University of Cologne, 50937 Cologne, Germany}

\begin{abstract}
It is a recent realization that many of the concepts and tools of causal discovery in machine learning are highly relevant to problems in quantum information, in particular quantum nonlocality. The crucial ingredient in the connection between both fields is the tool of Bayesian networks, a graphical model used to reason about probabilistic causation. Indeed, Bell's theorem concerns a particular kind of a Bayesian network and Bell inequalities are a special case of linear constraints following from such models. It is thus natural to look for generalized Bell scenarios involving more complex Bayesian networks. The problem, however, relies on the fact that such generalized scenarios are characterized by polynomial Bell inequalities and no current method is available to derive them beyond very simple cases. In this work, we make a significant step in that direction, providing a general and practical method for the derivation of polynomial Bell inequalities in a wide class of scenarios, applying it to a few cases of interest. We also show how our construction naturally gives rise to a notion of nonsignalling in generalized networks.
\end{abstract}

\maketitle

Bell's theorem \cite{Bell1964} demonstrates that our classical conceptions of causal relations must be taken with care, as they fail to commit with the results obtained in some quantum experiments performed by distant parties, the phenomenon known as quantum nonlocality. Even without detailed information about the underlying processes, the causal structure of the setup alone already implies strong constraints -- the famous Bell's inequalities -- on the correlations that are compatible with it.

This is close to the reasoning employed in the field of causal inference \cite{Pearlbook,Spirtesbook}, a connection that has recently attracted considerable attention \cite{Steeg2011,Fritz2012,Fritz2014,Chaves2014b,Spekkens2015,Henson2015,Chaves2015a,Chaves2015b,Ried2015}. Since Bell's theorem is a statement about classical correlations, it comes as no surprise that mathematical tools and concepts, originally devised in a causal inference context, can also be applied to the study of nonlocality. Indeed, Bell's theorem concerns the same kind of causal structures that are the object of study in Bayesian networks \cite{Pearlbook} and Bell inequalities are a special case of linear constraints following from such models \cite{Steeg2011}. Bayesian networks not only offer a new conceptual perspective to revisit quantum nonlocality \cite{Spekkens2015,Chaves2015b} but also provide the right language to devise generalized Bell scenarios \cite{Fritz2012,Fritz2014}.

Several extensions of the paradigmatic Bell experiment -- two distant parties, performing two possible experiments on their shares of a joint system -- have been proposed, including more parties
\cite{Werner2001,Zukowski2002}, more measurements/outcomes \cite{Collins2002,Collins2004} and sequential measurements \cite{Popescu1995,Gallego2014}. However, all these different generalizations share the same basic property: the correlations between all the parties originate from a single (not directly observable) source, being therefore named as local hidden variable (LHV) models. In spite of the rich plethora of phenomena and applications \cite{Brunner2014}, LHV models represent a very particular case of the possibilities offered by Bayesian networks. Those typically include several independent hidden variables and will be named here as generalized local hidden variable (GLHV) models. These scenarios with many independent sources are also ubiquitous in quantum information, e.g., entanglement percolation \cite{Acin2007}, entanglement swapping \cite{Zukowski1993} and quantum repeaters \cite{Sangouard2011,Sen2005}. Thus, understanding generalized Bell scenarios is not only of fundamental interest but also of high practical relevance.

Within that context, the basic question to be solved is how to derive Bell inequalities for general Bayesian networks. Bell inequalities play a fundamental role in study of nonlocality, since it is via their violation (e.g. with quantum entangled states) that we can witness the nonlocal character of a given experimental data. Unfortunately, as opposed to usual Bell scenarios, a GLHV model implies a non-convex region -- characterized by polynomial Bell inequalities -- of correlations that are compatible with it. Generally, algebraic geometry methods can be used to characterize such polynomial constraints \cite{Geiger1999,Garcia2005}, but given their computational complexity, in practice they are intractable even for very simple models \cite{Steeg2011}. Arguably, because of this difficulty, only sparse results have been obtained in the derivation of Bell inequalities for GLHV models, either using coarse-grained information \cite{Steudel2015,ChavesFritz2012,Fritz2012,Chaves2014a,Henson2015} or considering particular scenarios \cite{Branciard2010,Branciard2012,Tavakoli2014,Mukherjee2014}. However, to our knowledge, no practical and systematic method for the derivation of polynomial Bell inequalities for GLHV models is known to this date.

In this paper we propose a general method for deriving polynomial Bell inequalities in a wide class of Bayesian networks. In spite of the non-convex character of the problem, we show how to obtain polynomial inequalities resorting to a linear programming technique, namely a Fourier-Motzkin elimination \cite{Williams1986}. We illustrate the general method applying it to a few relevant cases and derive new polynomial Bell inequalities. Furthermore, we explain how our construction naturally leads to a notion of nonsignalling correlations \cite{Popescu1994} in generalized Bell networks.

\begin{figure}[t]
\includegraphics[width=0.95\columnwidth]{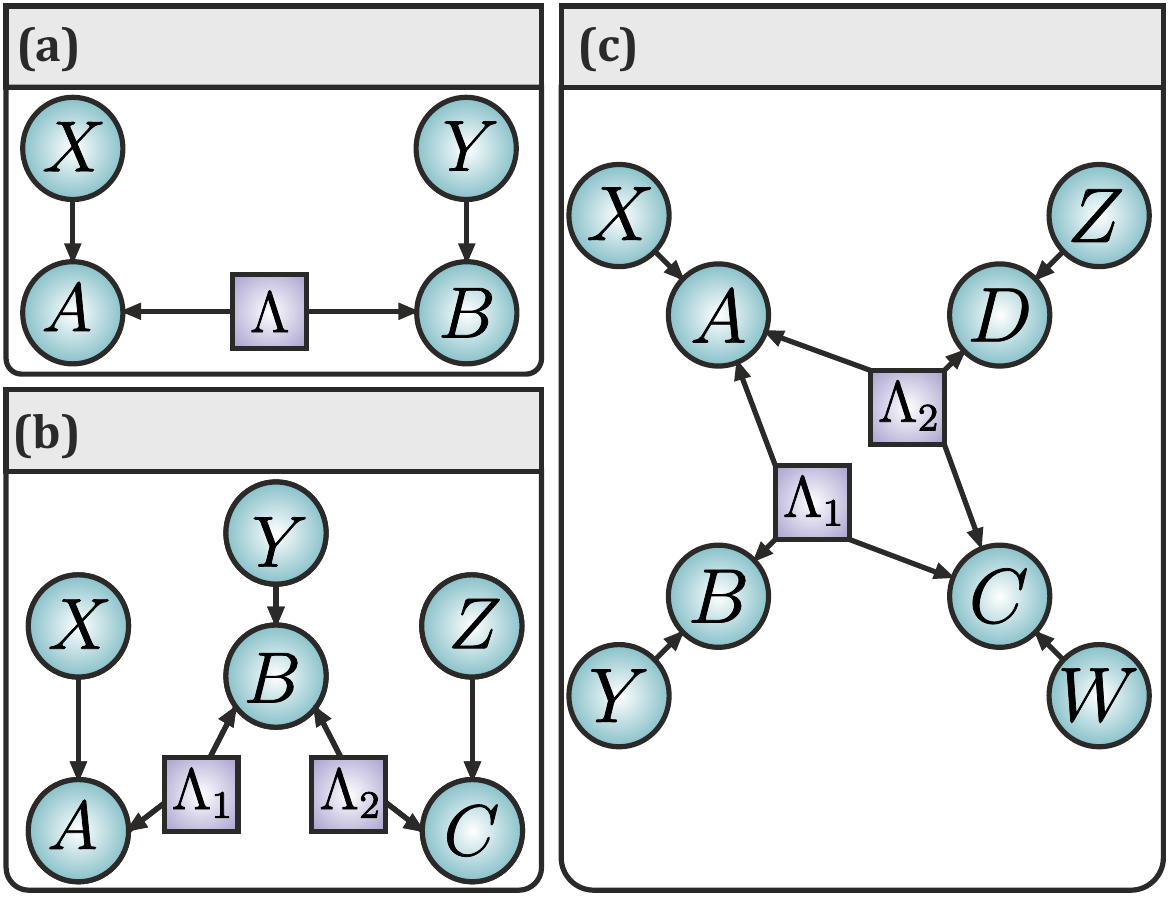}
\caption{DAG representation of Bayesian networks. \textbf{(a)} Bipartite LHV model. \textbf{(b)} GLHV model with $2$ independent hidden variables representing the bilocality scenario of \cite{Branciard2010}. \textbf{(c)} GLHV model with $2$ hidden variables shared among $4$ parties.}
\label{fig:DAGs}
\end{figure}

\section{Bell inequalities, Bayesian networks and marginal problems} Bell scenarios beyond LHV models can be represented via the graphical notation of Bayesian networks \cite{Pearlbook,Spirtesbook}. Underlying models are represented by directed acyclic graphs (DAG), where nodes stand for variables and directed arrows represent their causal relations \cite{Pearlbook}. While LHV models correspond to a DAG with a single hidden variable (see \figref{fig:DAGs}(a)), GLHV models are represented by DAGs with $n \geq 2$ independent hidden variables (see \figref{fig:DAGs}(b)-(c)). The causal relations implied by a DAG are captured by the (conditional) independencies (CI) implied by the graph and that can be listed by the d-separation criterion \cite{Pearlbook}. For instance, for the LHV model in \figref{fig:DAGs}(a) it follows that $p(x,y,\lambda)=p(x)p(y)p(\lambda)$ and $p(a\vert x,y,\lambda)=p(a\vert x,\lambda)$ (similarly to $b$). Thus, any observable data -- given by the probability distribution $p(a,b\vert x,y)$ --  compatible with this LHV model can be decomposed as
\begin{equation}
\label{LHV}
p(a,b\vert x,y)= \sum_{\lambda} p(a\vert x,\lambda)p(b\vert y,\lambda)p(\lambda).
\end{equation}
That is, any local distribution must lie inside the convex set defined by \eqref{LHV}, the so-called correlation polytope $\mathbb{C}$ \cite{Pitowsky1989,Pitowsky1991}. In this geometric picture, (linear) Bell inequalities are nothing else than facets of $\mathbb{C}$. Given that it is easy to list the extremal points of $\mathbb{C}$, to find its facets amounts to an efficient linear program, arguably the reason why this method has become the most prominent in the study of nonlocality.

Another equivalent, but far less used method, comes from the realization that Bell inequalities are constraints arising from a marginal problem \cite{Budroni2012,FritzChaves2013,Kaszlikowski2014}, that can be stated as: given some marginal distributions of $n$ variables is it possible to find a joint distribution of all variables, such that this distribution marginalizes to the given ones? To see that Bell's theorem is indeed a particular marginal problem, notice that the LHV description \eqref{LHV} is equivalent to the existence of a joint distribution $\p=p(a_0,\dots,a_{m_x},b_0,\dots,b_{m_y})$ (represented as a vector $\p$) describing the probability for outcomes of all possible measurements, where $a_i$ labels the outcome $a$ given that $x=i=\left\{0,\dots,m_x \right\}$ and similarly for $b$. Since $\p$ defines a valid probability, it is constrained by a set of linear inequalities $L \p \geq 0$ given by $p_i \geq 0$ (positivity) and $\sum_{i} p_i=1$ (normalization) defining the simplex polytope $\mathbb{P}$ \cite{boyd_convex_2009}. Given that at each round of the experiment only one $a_i$ and one $b_j$ can be measured simultaneously, $\p$ defines a non-observable quantity. However, the constraints on $\p$ will also imply constraints on the level of the observable distributions $p(a_i,b_j)$. These are exactly Bell inequalities, that in this picture can be understood as a condition for the marginal problem to have a positive answer. Thus, to obtain Bell inequalities in this picture, we have to eliminate from our description all non-observable terms. This is a achieved via a FM elimination \cite{Williams1986}, a standard algorithm for the elimination of variables from a system of inequalities.

For simplicity and without loss of generality, in the remaining of the paper we focus on dichotomic outcomes (e.g. $a_i=0,1$). It is then convenient to consider the equivalent description of the problem in terms of the correlation vector $\E$ with components given by expectation values, e.g.,  $\langle A_iB_j \rangle= \sum_{a_i,b_j} (-1)^{a_i+b_j}p(a_i,b_j)$. The vectors $\E$ and $\p$ are linearly related as $\E=T^{-1}\p$ implying that $\E$ must obey linear inequalities $T \E \geq 0$ plus a normalization constraint. To illustrate the FM elimination consider the CHSH scenario \cite{Clauser1969} where each of the two parties in \figref{fig:DAGs}(a) can measure two observables. The inequalities
\begin{eqnarray}
\label{eq:CHSH_almost_there1}
& & \langle A_0B_0 \rangle+\langle A_1B_0 \rangle-\langle A_0A_1 \rangle \leq 1, \\
\label{eq:CHSH_almost_there2}
& & \langle A_0B_1 \rangle-\langle A_1B_1 \rangle+\langle A_0A_1 \rangle \leq 1,
\end{eqnarray}
directly follow from $T \E \geq 0$ after the elimination of terms like $\langle A_0A_1B_0B_1 \rangle$ and $\langle A_0A_1B_0 \rangle$. The sum of \eqref{eq:CHSH_almost_there1} and \eqref{eq:CHSH_almost_there2} eliminates the remaining non-observable term $\langle A_0A_1 \rangle$ leading exactly to the CHSH inequality \cite{Clauser1969}.

\section{Polynomial Bell inequalities} Similarly to LHV models, a GLHV model also implies the existence of a well defined a joint distribution $\p$ characterized by linear inequalities $L \p \geq 0$. The difference resides on the fact that GLHV models also imply a set of non-linear inequalities $W \p \geq 0$ (where $W=W(\p)$). Thus, a GLHV model is characterized by intersection of $\mathbb{P}$ with $W \p \geq 0$, that is, a semi–algebraic set \cite{Geiger1999}. As discussed before, this system of inequalities involves non-observable quantities that have to be eliminated in order to obtain a description in terms of empirically accessible variables only. Formally, the problem at hand is equivalent to a quantifier elimination: the projection of a semi–algebraic set onto a subspace of it, that by Tarski-Seidenberg theorem is again guaranteed to be a semi–algebraic set \cite{Geiger1999}. In other terms, the correlations compatible with a GLHV model are characterized by finitely many polynomial Bell inequalities. Quantifier elimination is routinely encountered in algebraic geometry problems, thus general purpose methods have been developed \cite{Geiger1999}. Unfortunately, given their computational complexity, their application to Bell scenarios is intractable even for the simplest possible models \cite{Steeg2011}.
Notwithstanding, we show next that a simple adaptation of the FM elimination leads to practical and computational tractable way for deriving polynomial Bell inequalities.

The class of DAGs we consider are those which display (conditional) independencies on the level of the joint distribution $\p$.
This is the case, for instance, in the DAG of \figref{fig:DAGs}(b) implying the independence relation $p(a,c)=p(a)p(c)$ and for many other relevant scenarios in quantum information \cite{Zukowski1993,Sen2005,Acin2007,Branciard2010,Fritz2012,Tavakoli2014,Mukherjee2014}.
The method to derive polynomial inequalities for this class of scenarios proceed as follows.

Given $\p$ we first need to list all its components that are to be eliminated from our description: $\p_{\mathrm{O}}$ and $\p_{\mathrm{NO}}$ stand, respectively, to the set of observable and non-observable (to be eliminated) components $p_i$. We also list all the terms in $\p_{\mathrm{NO}}$ appearing in a non-linear fashion in $W \p \geq 0$, labeled by $\p_{\mathrm{WNO}}$. Notice that all terms appearing in $\p_{\mathrm{NO}}$ but not in $\p_{\mathrm{WNO}}$ can be eliminated via a usual FM elimination over $L \p \geq 0$, obtaining a new set of linear relations $L^{\prime} \p \geq 0$. The terms in $\p_{\mathrm{WNO}}$ have to be eliminated considering  $L^{\prime} \p \geq 0$ and $W \p \geq 0$ jointly. To that aim, notice that $W \p \geq 0$ can be linearized by considering some of the variables as free parameters of the problem.
Given $W \p \geq 0$ there is going to be a minimum set of variables $\p^{\prime}_{\mathrm{WNO}}$ that need to be set to free parameters in order to linearize the problem. This means that we can apply a FM elimination to the remaining terms obtaining a final set of inequalities that will depend linearly on the observable terms $\p_{\mathrm{O}}$ and polynomially on terms $\p^{\prime}_{\mathrm{WNO}}$. The observable data will also imply linear constraints on the parameters $\p^{\prime}_{\mathrm{WNO}}$. Together with these constraints, the obtained polynomials can be further simplified by usual quantifier elimination methods, finally arriving at polynomial inequalities involving observable data only. We highlight that following this procedure, one can derive all polynomial Bell inequalities following from the intersection of $L \p \geq 0$ and $W \p \geq 0$, that is, our method provides a full characterization of the GLHV models under consideration. In practice, however, a partial characterization (e.g. in terms of full correlators only) will often be the only computationally tractable approach.

To illustrate the general method, we start considering the bilocality scenario, one of few cases for which polynomial Bell inequalities are known \cite{Branciard2010,Branciard2012,Tavakoli2014}. The scenario involves three parties with correlations mediated via two independent sources (see \figref{fig:DAGs}(b)). In the particular case of two dichotomic measurements per party, the following inequality has been proven to hold \cite{Branciard2010,Branciard2012,Tavakoli2014}
\begin{equation}
\label{bilocality}
\sqrt{\vert I \vert}+\sqrt{\vert J \vert} \leq 2,
\end{equation}
where $I=\sum_{x,z=0,1}\mean{A_xB_0C_z}$ and $J=\sum_{x,z=0,1} (-1)^{x+z}\mean{A_xB_1C_z}$. However, the methods in \cite{Branciard2012,Tavakoli2014} cannot be easily generalized to different scenarios, for instance, considering three measurements per party. Next we show in details how our framework can be employed to easily prove \eqref{bilocality}. We then proceed to derive new polynomial Bell inequalities.

To derive \eqref{bilocality}, we consider the independence constraint following from the DAG in \figref{fig:DAGs}(b):
\begin{equation}
\label{bilocalE}
\mean{A_0A_1C_0C_1}=\mean{A_0A_1}\mean{C_0C_1}.
\end{equation}
We need to combine \eqref{bilocalE} via a FM elimination with the linear inequalities $T \E \geq0$. It is sufficient to consider two inequalities following from $T \E \geq0$:
\begin{eqnarray}
\label{TE1}
& & \pm I-\mean{A_0A_1}-\mean{C_0C_1}-\mean{A_0A_1C_0C_1} \leq 1, \\
\label{TE2}
& & \pm J+\mean{A_0A_1}+\mean{C_0C_1}-\mean{A_0A_1C_0C_1} \leq 1.
\end{eqnarray}
Substituting \eqref{bilocalE} in \eqref{TE1} and \eqref{TE2} and after some algebraic manipulations, we can combine both inequalities into a single polynomial inequality
\begin{equation}
\label{bineq_almost_there}
2\mean{A_0A_1}^2+(\pm J \mp I)\mean{A_0A_1}-(\pm I \pm J+2)\leq 0.
\end{equation}
As discussed before, we arrive at an inequality that depends linearly on the observable data (terms $I$ and $J$) but have a non-linear dependence on non-observable terms, in this case $\mean{A_0A_1}$. The minimum of the polynomial in \eqref{bineq_almost_there} is achieved at $\mean{A_0A_1}=(\pm I \mp J)/4$, implying the inequality in terms of observable data only
\begin{equation}
\label{eq:bilocalfinal}
-(1/8)(\pm I- \mp J)^2 -(\pm I \pm J+2) \leq 0.
\end{equation}
This is a quadratic equation that can be easily solved, e.g. for $I$, leading exactly to \eqref{bilocality}.

Another nice feature of our construction is the fact that independencies are not required to hold exactly. For instance, we may be interested in quantifying how much a given constraint must be relaxed in order to explain some experimental data \cite{Hall2010,Chaves2015b}. In the bilocality scenario if we allow for correlations $\mathcal{C}_{AC} \geq \vert \mean{A_0A_1C_0C_1}-\mean{A_0A_1}\mean{C_0C_1} \vert$ between parts A and C, it follows that
\begin{equation}
\label{eq:bilocalfinal_corr}
-(1/8)(\pm I- \mp J)^2 -(\pm I \pm J+2)  \leq 2\mathcal{C}_{AC},
\end{equation}
that is, the violation of \eqref{eq:bilocalfinal} quantifies the degree of correlation required to classically reproduce some non-bilocal correlation. As an illustration, consider the correlation $I=J=2$ that can be achieved quantum mechanically with two copies of Bell states shared between the parties \cite{Tavakoli2014}. In order to be classically reproduced, this correlation requires $\mathcal{C}_{AC}=1$, that is, maximal correlation between parts A and C.

To further illustrate the practicality and relevance of our method we also derived new polynomial inequalities. See the Appendix for a detailed discussion. For the considerably more complicated GLHV model in \figref{fig:DAGs}(c) the inequality \eqref{eq:bilocalfinal} is also valid if we define new functions given by $I=-\mean{A_1B_0C_0D_0}-\mean{A_1B_0C_0D_1}+\mean{A_1B_1C_0D_0}+\mean{A_1B_1C_0D_1}$ and $J=\mean{A_0B_0C_1D_0}-\mean{A_0B_0C_1D_1}+\mean{A_0B_1C_1D_0}-\mean{A_0B_1C_1D_1}$. Considering the bilocality scenario in \figref{fig:DAGs}(b) with $3$ measurement settings, the following inequality holds:
\begin{equation}
\label{eq:bilocal33}
-(1/8)(I-J+16)^2 + 8I \leq 0,
\end{equation}
with $I=\sum_{x,z=0,1,2}\mean{A_xB_0C_z}$ and $J=\sum_{x,z=0,1,2} (-1)^{x+z}\mean{A_xB_1C_z}$. To show the relevance of this inequality, notice that without the independence constraint it follows that $\vert I \vert +\vert J \vert \leq 10 $. Choosing a correlation given $I=J=9v$ (achievable in quantum mechanics for $v \leq 1/2$) we see that only for $v > 5/9$ the correlation is nonlocal. However, using \eqref{eq:bilocal33} we see that this correlation is non-bilocal for $v > 4/9$, illustrating the gap between the local and bilocal sets.

\section{Nonsignalling correlations and generalized Bayesian networks} In the study of nonlocality it is often useful to define the notion of nonsignalling (NS) correlations \cite{Barrett2005}. These are the observable distributions $\p_{\mathrm{O}}$ that cannot be used to signal between the parties, that is, the marginal distributions are well defined quantities that cannot depend in any way on which observable the other parties have measured. A paradigmatic example of a NS correlation is the Popescu-Rohrlich(PR)-box defined as $p(a,b \vert x,y) = 1/2 \delta_{a\oplus b,xy}$. \cite{Popescu1994}.

The marginal problem approach naturally incorporates the notion of NS correlations \cite{Popescu1994}. For instance, for the bipartite scenario in \figref{fig:DAGs}(a), NS correlations are those that have well defined observable distributions $p(a_i,b_j) \forall i,j$ (respecting positivity and normalization) and well defined marginals, that is, $p(a_i)=\sum_{b_j} p(a_i,b_j)=\sum_{b_{j^{\prime}}} p(a_i,b_{j^{\prime}})$ $\forall j,j^{\prime}$ (and similarly for $p(b_j)$). These constraints can be combined into a system of linear inequalities $L_{\mathrm{NS}} \p_{\mathrm{O}} \geq 0$ defining a polytope that is characterized by finitely many extremal points.

We can define NS correlations in generalized Bayesian networks, as those that are compatible with GLHV models where all the underlying (classical) hidden variables are replaced by general NS distributions. As shown in \cite{Henson2015}, all the conditional independencies on the level of the observable distributions $\p_{\mathrm{O}}$ that are valid in the classical setup will remain valid after this replacement. For instance, for the bilocality scenario in \figref{fig:DAGs}(b), even if we allow for NS correlations to be shared between the parties, it is true that the statistics between parties A and C should factorize, that is, $p(a_i,c_k)=p(a_i)p(c_k)$. As before, we can represent the observable independencies as a system of polynomial inequalities $W_{\mathrm{NS}} \p_{\mathrm{O}} \geq 0$. Thus, we can define generalized nonsignalling (GNS) correlations as those inside the semi-algebraic set $\Sigma$, defined by intersection of $L_{\mathrm{NS}} \p_{\mathrm{O}} \geq 0$ and $W_{\mathrm{NS}} \p_{\mathrm{O}} \geq 0$. Since $\Sigma$ defines a non-convex body characterized by polynomial inequalities, differently from the usual case, there are going to be infinitely many extremal GNS points defining it. In spite of that, we can still define a sensible and practical way to characterize GNS correlations. Similarly to what has been done before, we can take some of the variables appearing in $W_{\mathrm{NS}} \p_{\mathrm{O}} \geq 0$ as free parameters in order to linearize it. Doing that we turn $\Sigma$ into a convex set with finitely many extremal points that can therefore be characterized by standard linear program techniques.

As an illustration consider the GLHV model in \figref{fig:DAGs}(b) with all parties performing two dichotomic measurements. If we fix the marginal distributions of parts A and C to be $p(a_i)=p(c_k)=1/2$, we see that one of the extremal GNS points is given by $p(a,b,c \vert x,y,z) = 1/4 \delta_{a\oplus b\oplus c,y(x \oplus z)}$, a distribution that can be achieved replacing the hidden variables in \figref{fig:DAGs}(b) by PR-boxes \cite{Barrett2005}.

\section{Discussion} Bayesian networks offer an almost unexplored ground for generalizations of Bell's theorem. The basic question to be solved in this quest is how to derive polynomial Bell inequalities associated with more complex causal structures. In this work we made an important step in that direction. We proposed a practical and general method that can be readily applied to a wide range of scenarios, considering its applications in few GLHV models and deriving polynomial Bell inequalities characterizing them. We have also shown how our construction naturally leads to a notion of nonsignalling correlations in GLHV models.

Given the fundamental role that Bell inequalities play in the study and practical applications of nonlocality, we believe that our results will motivate and set a basic tool for future research in generalized Bell scenarios. The natural next step is to put the machinery to use in a variety of scenarios and derive new Bell inequalities well suited, for example, to decrease the requirements on experimental implementations of Bell tests \cite{Eberhard1993}. It would be interesting to investigate the role of polynomial Bell inequalities in practical applications of nonlocality, such as quantum cryptography \cite{Acin2007b}, randomness generation \cite{Colbeck2007,Pironio2010} or distributed computing \cite{Brukner2004}. For instance, the amount of violation of usual Bell inequalities can be directly associated with the probability of success in communication complexity problems \cite{Brukner2004,Buhrman2010}. Are there any communication problems associated to polynomial Bell inequalities? Another possibility is to find Tsirelson's bounds \cite{Tsirelson1980,Navascues2007} associated with these generalized inequalities, that is, what is the maximum violation of them achievable with quantum correlations. Related to that and inspired by results such as information causality \cite{Pawlowski2009}, it would also be relevant to derive information-theoretical principles for these more complex Bayesian networks \cite{Chaves2015b}.

\begin{acknowledgments}
We acknowledge financial support from the Excellence Initiative of the German Federal and State Governments (Grants ZUK 43 \& 81), the US Army Research Office under contracts W911NF-14-1-0098 and W911NF-14-1-0133 (Quantum Characterization, Verification, and Validation), the DFG (GRO 4334 \& SPP 1798).
\end{acknowledgments}

\bibliography{NLinearBell}

\appendix

\begin{widetext}

\section{A method for the derivation of polynomial Bell inequalities}
As discussed in the main text, the derivation of polynomial Bell inequalities follows from an adapted FM elimination over the combined system of inequalities $L \p \geq 0$ and $W \p \geq 0$ (with $W=W(\p)$), the first representing linear relations respected by a well defined probability distribution $\p$ while the latter stands for the (conditional) independence (CI) constraints implied by a given GLHV model. Notice, however, that not all DAGs will display CIs on the level of $\p$; this is the case for instance in the so-called triangle scenario \cite{Steudel2015,Fritz2012,Chaves2014a}.

Given the scenario of interest, we need to define $\p_{\mathrm{O}}$ and $\p_{\mathrm{NO}}$ standing, respectively, to the set of components $p_i$ that we want to keep or not in our description. We also need to define $\p_{\mathrm{WNO}}$ and $\p^{\prime}_{\mathrm{WNO}}$. The first corresponds to the components in $\p_{\mathrm{NO}}$ appearing in a non-linear fashion in the inequalities $W \p \geq 0$ while the latter describes the minimum set of components that need to be taken as free real parameters in order to linearize $W \p \geq 0$. All the terms in $\p_{\mathrm{NO}}$ but not in $\p_{\mathrm{WNO}}$ can be eliminated via a usual FM elimination leading to new set of inequalities $L^{\prime} \p \geq 0$. To understand the FM elimination, notice that since the sum of two valid inequalities also defines a valid inequality, in order to eliminate a given term from our description we basically have to consider all possible pairwise sums of inequalities where the coefficients of the term to be eliminated appear with opposite signs. The remaining terms have to be eliminated resorting to the adapted FM method discussed in the main text and illustrated in details below.

As a side remark, we notice that instead of performing the usual FM elimination leading to $L^{\prime} \p \geq 0$, one can equivalently list the extremal points in the subspace given by the support of $L^{\prime}$, and then dualize the description in order to exactly obtain the inequalities $L^{\prime} \p \geq 0$. In practice which approach will be better is going to depend on the scenario in question. Typically, if the number of terms in $\p_{\mathrm{O}}$ is not large, the dualization approach will be reasonably faster. We refer the reader to Ref. \cite{Budroni2012} for a discussion of the computational advantage of the both methods in usual LHV models. In the following, for simplicity and without loss of generality, we focus on the case where all measurements have dichotomic outcomes so that we can equivalently treat the problem in terms of the correlation vector $\E$ with components given by expectation values.

To illustrate the abstract discussion, consider the bilocality scenario (see Fig. 1(b) in the main text) in the particular case where all the parties measure two dichotomic observables, implying the bilocality constraint $p_{ac}(a_0,a_1,c_0,c_1)=p_{a}(a_0,a_1)p_{c}(c_0,c_1)$. Since the variables are binary, we see that the bilocality assumption is equivalent to $16$ (not necessarily independent) quadratic constraints. In order to linearize this set of constraints we can take $v_{a_0,a_1}=p_{a}(a_0,a_1)$ (for each $a_0,a_1=0,1$) as a free real parameter, that is, we can express the non-linear constraints $W(\p) \p \geq 0$ as a linear relation  $W(v_{0,0},v_{0,1},v_{1,0},v_{1,1}) \p \geq 0$. In terms of expectation values, we have a correlation vector $\E=(1,E_{C_0},E_{C_1},E_{C_0C_1}, \dots, E_{A_0A_1B_0B_1C_0C_1})$ (for simplicity we label $\mean{X}=E_{X}$) with $64$ components that must respect the linear constraints $T \E \geq 0$ (with $\E=T^{-1}\p$). The bilocality constraints can also be expressed in terms of expectation values. For instance, $p_{ac}(0,0,0,0)=p_{a}(0,0)p_{c}(0,0)$ is equivalent to
\begin{eqnarray}
\label{bilocalE1}
& & v (1 + E_{C_1}+  E_{C_0}+ E_{C_0C_1}) + E_{A_0A_1} + E_{A_0}+ E_{A_1}+ E_{A_0C_0}+ E_{A_0C_1}+ E_{A_1C_0}+ E_{A_1C_1} \\ \nonumber
& &+ E_{A_1C_0C_1}+E_{A_0C_0C_1}+ E_{A_0A_1C_1}+ E_{A_0A_1C_0}+ E_{A_0A_1C_0C_1} =0
\end{eqnarray}
with $v=1-4p_{a}(0,0)$, where $p_{a}(0,0)$ is the real free parameter.
Notice that the bilocality constraints do not depend on the variables $B_0$ and $B_1$. Therefore all non-observable terms that depend on them, for instance, $E_{A_0A_1B_0B_1C_0C_1}$, can be eliminated via the usual FM elimination method over $T \E \geq 0$, defining a new system of linear inequalities $T^{\prime} \E \geq 0$. The remaining terms to be eliminated are those that depend jointly on $A_0,A_1$ and/or $C_0,C_1$. We further notice that all the bilocal constraints (e.g. \eqref{bilocalE1}) only have a non-linear dependence on the terms $E_{C_0}$, $E_{C_1}$ and $E_{C_0C_1}$. That is, all the non-observable terms but $E_{C_0C_1}$ can be eliminated via the usual FM elimination. After all non-observable terms have been eliminated we arrive at a final description that depends linearly on the observable terms and non-linearly on the free parameters $v_{a_0,a_1}$. Notice that the observable data will also imply linear constraints on the free parameters themselves. Together with these constraints, each of obtained polynomial inequalities can be further simplified by usual quantifier elimination methods (see for instance the function \emph{Reduce} in Mathematica \cite{Mathematica}), finally arising at polynomial inequalities involving the observable data only.

In the following we will apply the general method to each of the scenarios in Fig. 1 of the main text. For computational reasons we consider the case with full correlators but no marginal terms, that is, we keep terms like $\mean{A_iB_jC_k}$ but not terms like $\mean{A_iB_j}$ or $\mean{A_i}$. Notwithstanding, as highlighted in the main text, our method can also be applied to obtain a full characterization of the GLHV models, that is, including marginal terms.

\section{Detailed derivation of the polynomial Bell inequalities}
We start considering the bilocality scenario in Fig. 1(b) in the case where each party A,B and C can measure two dichotomic observables. We highlight that the same analysis remains valid if we consider part B to perform a single measurement with $4$ possible outcomes, e.g., a measurement in the Bell basis. Restricting to the subspace of full correlators --that is, containing terms $\mean{A_iB_jC_k}$ -- we observe that one of the obtained inequalities is exactly eq. (8) of the main text. That is, this inequality corresponds to a facet of the bilocal set in the subspace of full correlators. We further notice that in order to derive this class of inequalities it is sufficient to consider -- instead of the full set of bilocal constraints --  the simple constraint
\begin{equation}
\label{bilocalE_app}
\mean{A_0A_1C_0C_1}=\mean{A_0A_1}\mean{C_0C_1}.
\end{equation}
Together with the linear constraints
\begin{eqnarray}
\label{TEa_app}
& & \pm I-\mean{A_0A_1}-\mean{C_0C_1}-\mean{A_0A_1C_0C_1} \leq 1, \\
\label{TEb_app}
& & \pm J+\mean{A_0A_1}+\mean{C_0C_1}-\mean{A_0A_1C_0C_1} \leq 1,
\end{eqnarray}
we can readily prove inequality  eq. (8) of the main text. Inequalities \eqref{TEa_app} and \eqref{TEb_app} directly follow from $T \E \geq 0$ with $I=\sum_{x,z}\mean{A_xB_0C_z}$ and $J=\sum_{x,z} (-1)^{x+z}\mean{A_xB_1C_z}$.  Substituting \eqref{bilocalE_app} in \eqref{TEa_app} and \eqref{TEb_app} we obtain
\begin{eqnarray}
\label{TE3a_app}
& & \pm I-(1+\mean{A_0A_1})-\mean{C_0C_1}(1+\mean{A_0A_1}) \leq 0, \\
\label{TE3b_app}
& & \pm J-(1-\mean{A_0A_1})+\mean{C_0C_1}(1-\mean{A_0A_1}) \leq 0.
\end{eqnarray}
Notice that  $(1 \pm \mean{A_0A_1}) \geq 0$, with equality only if $\mean{A_0A_1}=\mp 1$, that is, only if both outputs of part A are deterministic functions, that is, only if we have the trivial case where either $I=0$ or $J=0$. For $(1\pm \mean{A_0A_1}) > 0$, we can rearrange \eqref{TE3a_app} and \eqref{TE3b_app} as
\begin{eqnarray}
\label{TE4_app}
& & \pm I/(1+\mean{A_0A_1})-1-\mean{C_0C_1} \leq 0, \\
& & \pm J/(1-\mean{A_0A_1})-1+\mean{C_0C_1} \leq 0.
\end{eqnarray}
Summing both inequalities we eliminate the term $\mean{C_0C_1}$ and arrive at
\begin{equation}
\label{bineq_almost_there_app}
\pm I/(1+\mean{A_0A_1}) \pm J/(1-\mean{A_0A_1})\leq 2,
\end{equation}
that can be further arranged to obtain the class of inequalities discussed in the main text, given by
\begin{equation}
\label{bineq_almost_there_app2}
2\mean{A_0A_1}^2+(\pm J \mp I)\mean{A_0A_1}-(2 \pm I \pm J)\leq 0.
\end{equation}
As discussed before, we arrive at an inequality that depends linearly on the observable data (terms $I$ and $J$) but have a non-linear dependence on the non-observable term $\mean{A_0A_1}$. Given $I$ and $J$, to check that this data fulfills the inequality, we have to prove that there is at least one choice of $\mean{A_0A_1}$ such that the lhs of \eqref{bineq_almost_there_app2} is $\leq 0$. That is, to make use of \eqref{bineq_almost_there_app2} we have to find the value of $\mean{A_0A_1}$ (as a function of $I$ and $J$) minimizing the polynomial on the lhs. Since the lhs in \eqref{bineq_almost_there_app2} defines a convex function, the minimum of the inequality is achieved at $\mean{A_0A_1}=(\pm I \mp J)/4$, implying the inequality in terms of observable data only.
\begin{equation}
\label{bineq_final_app}
-(1/8)(\pm I \mp J)^2 -(2\pm I \pm J) \leq 0.
\end{equation}

A similar derivation is possible if we allow for correlation between parts A and C, such that $\vert \mean{A_0A_1C_0C_1}-\mean{A_0A_1}\mean{C_0C_1} \vert  \leq \mathcal{C}_{AC}$. Summing $ \mean{A_0A_1C_0C_1}-\mean{A_0A_1}\mean{C_0C_1}  \leq \mathcal{C}_{AC}$ with \eqref{TEa_app} and \eqref{TEb_app} we obtain
\begin{eqnarray}
\label{TE3ac_app}
& & \pm I-(1+\mean{A_0A_1})-\mean{C_0C_1}(1+\mean{A_0A_1}) \leq \mathcal{C}_{AC}, \\
\label{TE3bc_app}
& & \pm J-(1-\mean{A_0A_1})+\mean{C_0C_1}(1-\mean{A_0A_1}) \leq \mathcal{C}_{AC}.
\end{eqnarray}
Proceeding with the exact same steps as above we finally obtain
\begin{equation}
\label{bineqc_final_app}
-(1/8)(\pm I \mp J)^2 -(2\pm I \pm J) \leq 2 \mathcal{C}_{AC}.
\end{equation}

To prove that a similar inequality holds for the GLHV model in Fig. 1(c) we can follow a very similar derivation. For this model it follows the independence constraint
\begin{equation}
\mean{B_0B_1D_0D_1}=\mean{B_0B_1}\mean{D_0D_1}.
\end{equation}
Together with the linear constraints
\begin{eqnarray}
& & \pm I-\mean{B_0B_1}+\mean{D_0D_1}+\mean{B_0B_1D_0D_1} \leq 1, \\
& & \pm J+\mean{B_0B_1}-\mean{D_0D_1}+\mean{B_0B_1D_0D_1} \leq 1,
\end{eqnarray}
where $I=-\mean{A_1B_0C_0D_0}-\mean{A_1B_0C_0D_1}+\mean{A_1B_1C_0D_0}+\mean{A_1B_1C_0D_1}$ and $J=+\mean{A_0B_0C_1D_0}-\mean{A_0B_0C_1D_1}+\mean{A_0B_1C_1D_0}-\mean{A_0B_1C_1D_1}$, we can follow the same steps as above to prove that \eqref{bineq_final_app} also holds in this scenario.

We now move to the scenario in Fig. 1(b) where parties A and C can measure three possible observables. To prove that inequality (11) of the main text holds in this case, we need to proceed as follows. We have to consider the inequalities
\begin{eqnarray}
\label{bi331_app}
& & 4J-3\mean{f_1}+\mean{f_1C_0C_1}-\mean{f_1C_0C_2}+\mean{f_1C_1C_2} \leq 0, \\
\label{bi332_app}
& & 4I-3\mean{f_2}-\mean{f_2C_0C_1}-\mean{f_2C_0C_2}-\mean{f_2C_1C_2} \leq 0,
\end{eqnarray}
that follow from $T \E \geq 0$ with $f_1=+3-A_0A_1+A_0A_2-A_1A_2$, $f_2=3+A_0A_1+A_0A_2+A_1A_2$. It also follows from $T \E \geq 0$ that
\begin{eqnarray}
\label{LHV331}
& & 3 f_2 \geq  2\vert I \vert , \\
\label{LHV332}
& & 3 f_1 \geq  2\vert J \vert, \\
\label{LHV333}
& & f_1 + f_2 \leq 8.
\end{eqnarray}

Using the independence relation
\begin{eqnarray}
& & \mean{A_iA_jC_kC_l}= \mean{A_iA_j}\mean{C_kC_l} \forall i,j,k,l
\end{eqnarray}
we can rewrite \eqref{bi331_app} and \eqref{bi332_app} as
\begin{eqnarray}
& & 4J+\mean{f_1}(-3+\mean{C_0C_1}-\mean{C_0C_2}+\mean{C_1C_2}) \leq 0, \\
& & 4I+\mean{f_2}(-3-\mean{C_0C_1}-\mean{C_0C_2}-\mean{C_1C_2}) \leq 0.
\end{eqnarray}

Since \eqref{LHV331} and \eqref{LHV332} imply that $f_1$ and $f_2$ are strictly positive quantifies (apart from the trivial case $I=0$ and/or $J=0$), we can rewrite these inequalities as
\begin{eqnarray}
& & 4J/{f_1}+(-3+\mean{C_0C_1}-\mean{C_0C_2}+\mean{C_1C_2}) \leq 0, \\
& & 4I/{f_2}+(-3-\mean{C_0C_1}-\mean{C_0C_2}-\mean{C_1C_2}) \leq 0. \\
\end{eqnarray}
Summing both inequalities we eliminate the terms $\mean{C_0C_1}$ and $\mean{C_1C_2}$, obtaining
\begin{equation}
4I/{f_2}+4J/{f_1}-2\mean{C_0C_2}) \leq 6.
\end{equation}
Combining it with the trivial inequality $\mean{C_0C_2} \leq 1$, we finally obtain
\begin{equation}
If_1+Jf_2 \leq 2f_1f_2.
\end{equation}

Since \eqref{LHV331} implies that $\pm I-2f_2$ is a strictly negative quantity (apart from the trivial case $\vert I \vert=2f_2$), we can rewrite the inequality above as
\begin{equation}
-f_1 - Jf_2/(I-2f_2) \leq 0.
\end{equation}
Summing it with \eqref{LHV331}, we obtain
\begin{equation}
f_2 - Jf_2/( I-2f_2) \leq 8,
\end{equation}
that can be rewritten as
\begin{equation}
2f_2^2+f_2(-I+J-16) + 8I \leq 0.
\end{equation}
The lhs is a quadratic equation on the non-observable term that is convex, implying that the minimum of the lhs is obtained at $f_2=(+ I - J+16)/4$ and therefore
\begin{equation}
-(1/8)(+ I - J+16)^2 + 8I \leq 0.
\end{equation}

\end{widetext}

\end{document}